# Microcontroller Based Testing of Digital IP-Core

Amandeep Singh[1] and Balwinder Singh[2]

[1-2]Acadmic and Consultancy Services Division,
Centre for Development of Advanced Computing(C-DAC), Mohali, India
`eramansingh@yahoo.com` , `balwinder_cdacmohali@yahoo.com`

## ABSTRACT

*Testing core based System on Chip [1] is a challenge for the test engineers. To test the complete SOC at one time with maximum fault coverage, test engineers prefer to test each IP-core separately. At speed testing using external testers is more expensive because of gigahertz processor. The purpose of this paper is to develop cost efficient and flexible test methodology for testing digital IP-cores [2]. The prominent feature of the approach is to use microcontroller to test IP-core. The novel feature is that there is no need of test pattern generator and output response analyzer as microcontroller performs the function of both. This approach has various advantages such as at speed testing, low cost, less area overhead and greater flexibility since most of the testing process is based on software.*

## 1. INTRODUCTION

Deep Sub-Micron (DSM) technology makes it feasible to integrate large number of blocks called Intellectual Property Core on a single System-on-Chip (SOC). Testing core based System on Chip [1]is a big challengefor test engineers. The reason is that it is impossible to test the System on Chip with full fault coverage. The novel approach is to test the each IP-core separately. IP-core consists of microprocessors (MP), microcontroller (MC), memories, ASICs, and peripherals. Here digital IP-core is used for testing. Digital IP-cores are also known as soft cores because these are synthesizable cores written in Hardware Description Language such as Verilog or VHDL. Soft cores allow synthesis, placement and route (SPR) design flow. Testing is necessary for production process in the VLSI technology, since it ensures the proper functioning of the design. Testing is done to detect the faults in case of faulty design. In VLSI [3], a design is tested by taking specifications into consideration such that the design is responding according to its specifications.

The basic principle consists of applying test vectors to input of the circuit under test and the output response is compared with the reference signature. If output matches, then the circuit is fault free, otherwise faulty. The same principle is used in our testing methodology.

The traditional approach is to test the design after completion but modern approach is to test the design at an early stage i.e. at design specifications level. The later has an advantage over the former as it is cost effective and faults are removed at design stage whereas in the former it is not possible to do so.FPGA, also called VLSI breadboard, is chosen for implementing IP-Core, providing flexibility in the design. Working prototype of microprocessor is developed in [4]. The FPGA technology is commonly used for circuit emulation. In our design embedded controller is implemented on FPGA for testing. Different types of testers are used for testing IP-cores. Dedicated testers are one of them which are used to perform at speed testing. These testers are very expensive because of gigahertz processor. The main elements of testers are test pattern generator and output response analyzer. In our approach microcontroller will perform the function of both. Hence our approach is cost efficient, as there is no need for expensive testers. Generating test sequence is the main problem. Test sequence generation for controller test necessarily requires the knowledge of controller instruction set and instruction format. Testing approach based on hardware redundancy technique–Triple Modular Redundancy (TMR)–for the most important MC's component like ALU and control unit is used. When a certain fault in one of the redundancy modules by the TMR voter is detected, a diagnosis approach is used to locate the faulty elements inside that faulty module by testing faulty module independently as previously done. We propose a flexible testing methodology to test the digital IP-core by using FPGA and microcontroller. Both provide flexibility as depend on the software. The testing is done by ATMEL ATmega8515 microcontroller by interfacing with FPGA on which the IP-core is implemented. The testing time is calculated by calculating the number of instructions in testing program used to test the core.

The paper is organized as follows. In section 2 theoutlines of related work are given. Section 3 describes theTest evaluation Frameworkalong with IP-core design and implementation and test methodology. Experimental Results are described in section 4.Section 5 shows the conclusion of the work.

## 2. RELATED WORK

More research work has been carried out in Testing of IP-Core. Some of them are discussed as: C. A. Papachristou [5] proposed flexible design for test methodology for testing a core-based system on chip (SOC). In this microprocessor/memory pair is embedded to test the remaining components of the SOC. In [6] a method for the generation of effective programs for the self-test of a processor is described. Here ATE is loaded into the memory and test program is generated and processor core can execute it whenever necessary. FPGA-based fault simulator is described in [7] where reconfiguration is performed by anembedded processor core. The advantage of the proposed technique is the reduction of reconfiguration time. Software fault simulation is described in [8]. The technique provides flexibility since it is VHDL software based and simulations are performed by the VHDL Simulators. The drawback of Software based fault simulation is that it is very computationally extensive. Another technique is to use Hardware based fault simulation which is faster than software based. Hardware part is emulated on FPGA. [9] Proposes an on-line testing approach based on hardware redundancy technique–triple modular redundancy (TMR). Software based self test methodology is proposed in [10], [11], [12].

Several techniques for testing involve Test Pattern Generator and Output Response analyzer separately and microprocessor core as a test controller. These techniques have several advantages such as at speed testing, less area overhead, flexibility. The test pattern generation is main problem as test patterns with high fault coverage are required. Methods to generate pseudo-exhaustive test sets at function level are given in [13]. The proposed method can be used to generate test sets for IP cores, specially for soft cores. Method for generation of effective test programs is described in [6].

## 3. TEST EVALUATION FRAMEWORK

At first test evaluation framework is developed, on the basis of which testing is performed. Framework helps in designing the IP-core, to generate the test program, to decide the test vectors to apply and lastly, the testing using microcontroller and FPGA. Test evaluation framework is explained as: VHDL simulation is performed by using RTL descriptions, from the simulations inputs are captured which are given to the IP-core i.e. test vectors. The assembler takes the test program written in assembly along with test vectors and loads it into the program memory of the microcontroller (ATmega8515). The test patterns are generated through the simulation as the inputs which are responsible for the functional outputs are taken. The microcontroller applies test pattern to FPGA and gets the response, compares with signature to indicate error if any. The novel approach is that there is no need of the Test Pattern generator and output response analyzer as microcontroller performs the function of both. Test evaluation framework is shown in Figure.1.

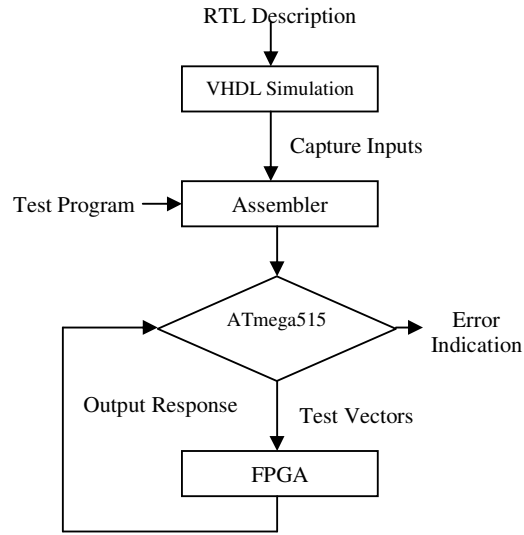

Figure.1. Test Evaluation Framework

### 3.1. Embedded Controller design and implementation

In this section the designing of IP-core is described, which is going to be tested. Here the embedded controller is designed as digital IP-core. The IP-Core is designed in VHDL and then implemented on FPGA. The embedded controller designed consists of the following basic blocks: Arithmetic and Logic Unit (ALU), Control Unit, Memory, Program Counter, Instruction Register as shown in figure 2. The various blocks are explained as:

*Program Counter:* In PC two control signals are required for the correct operation. The first is the signal to increment the PC (PC_inc), the second is the control signal load the PC with a specified value (PC_load).

*Arithmetic and Logic Unit (ALU):* The arithmetic and logic unit (ALU) has the same clock and reset signals as the PC, and also the same interface to the bus (ALU_bus) defined as type inout. The ALU also has two further control signals, ALU_valid and ALU_cmd which can be decoded to map to the seven individual functions required of the ALU.

*Instruction Register:* The IR also has two further control signals, the first being the command to load the IR (IR_load), and the second being to load the IR_Bus with the IR_internal (IR_valid). The final connection is the decoded opcode that is to be sent to the system controller.

*Memory (RAM):* The controller requires a RAM memory, with an address register (MAR) and a data register (MDR). Therefore needs to be a load signal for each of these registers: MDR_load_RAM and MAR_load_RAM. As it is a memory, there also needs to be an enable signal (M_en_ram), and also a signal denotes Read or Write modes (M_rw). We have designed 8x8 RAM.

*Memory (ROM):* The controller requires a ROM memory to hold the program to be executed. There is no read/write signal needed here. We have designed 8x16 ROM.

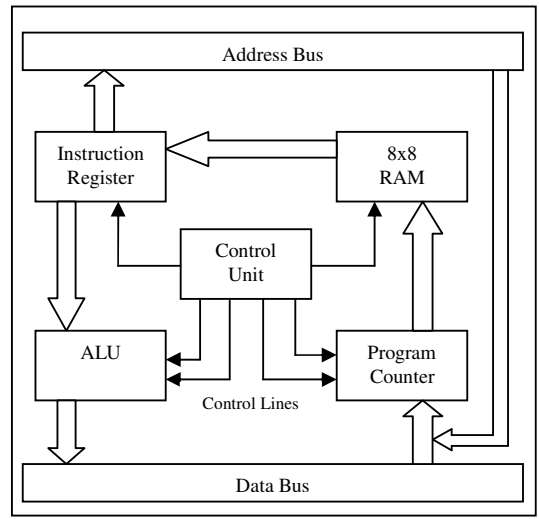

Figure.2. Controller Block Diagram

*Control Unit:* The control unit must have the clock and reset signals, a connection to the global bus and finally all the relevant control signals must be output.

### 3.2. Testing Methodology

Testing Methodology tells us the testing process. Before testing, one should know the specification of the design being tested [14]. The basic method of testing is applying binary patterns (test vectors) at input and output patterns (response) are compared with signature. Signature is the expected or correct response of the fault free circuit. If the response of the circuit under test matches with the signature, then the circuit is considered as fault free. The same method is used here to test the Digital IP-core.

ATMEL ATmega8515 microcontroller is used for testing. Since it is 8-bit microcontroller, it can be used to test 8-bit IP-core. In earlier section we have designed an 8-bit embedded controller and in this section we perform its testing. Before testing the embedded controller by embedding all the basic blocks, we should perform their testing independently to ensure the good fault coverage. The testing programs are written in assembly language of the microcontroller called macro for all the basic blocks. Hence the blocks are tested using macros. Test Vectors are applied from the microcontroller to the design implemented on FPGA and the response of the design is given back to the microcontroller again. The response of the design is compared with the signature that is stored in microcontroller and design is tested based on the comparison of the signature and the response. The work of test vectors generation, response analyzer and controlling the testing, all is performed by the microcontroller. Test Vectors are generated in pseudo random fashion through simulation as only that test vectors are considered which are responsible for the output response. The pseudo code for testing is written as:

ldi r16, (test vector)
out portd, r16

```
in r17, pinb
cpi r17, (signature)
brne error
```

As shown above in pseudo code one port is assigned as output port to generate test vectors and other port is assigned as input port to take the response from the design. After that the response is compared with signature, if it is matched then the next test vector is applied for further testing, if not then it gives error by branching to error subroutine which indicates error either by glowing LED or displaying on the LCD. Note that the next pattern is applied only if the response of the previous one is matched. Hence ensure the proper fault coverage, as it tells where the fault is present. The novelty of the testing is proper fault coverage by applying test vector one after other. After testing the blocks independently, embedded controller is tested by the same method of testing.

### 3.3. Triple Modular Redundancy Implementation

In order to make the design redundant and to mask the fault Triple Modular Redundancy (TMR) is used. TMR makes the design redundant and the faults,if any present in the design, are masked. In TMR three copies of the design is used and bit wise majority vote is performed on the output of the triplicate circuit as shown in Figure.3.

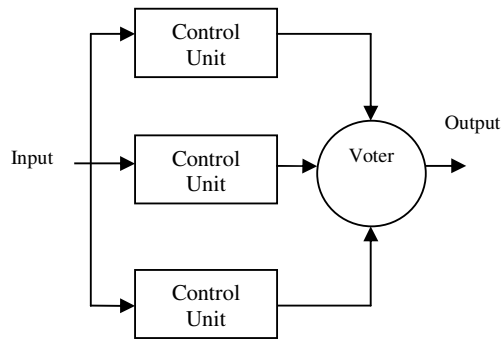

Figure.3. TMR for Control Unit

In our design the TMR is implemented on the most important parts i.e. ALU and Control Unit. As shown in figure three copies of the Control Unit are used. The input to the three Control units is same and the outputs of the Control units are fed to the majority voter unit. The function of the majority voter is to give the output that corresponds to at least two of its inputs. Hence fault if any present in the Control unit, it will affect the output since the other two will mask its fault called fault masking. The only limitation here is that if the voter is faulty then it would reflect in the whole design. Here we assume that the voter is tested and fault free. TMR will mask the faults present due to Single event upset (SEU), Stuck at fault. Diagnosis technique is used in parallel with TMR to locate the faulty elements. In diagnosis technique output of all three Control units are compared with the voter output, if there is any difference in the input and output, the faulty element will get detected. After that the faulty element will be tested by using macros that will locate fault in the unit. The testing platform is as shown in Figure .4, the microcontroller on AVR STK500 development Board is interfaced with the FPGA Spartan-3 development board on which embedded controller IP-Core is implemented.

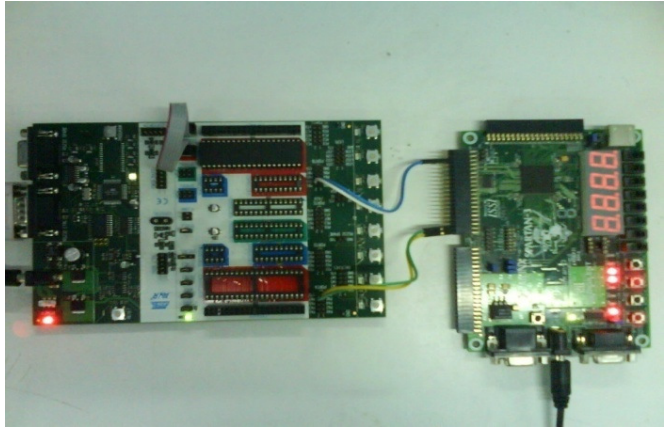

Figure.4. Testing Platform

## 4. EXPERIMENTAL RESULTS

Testing of the various modules is performed by applying the test vectors which are responsible for their operations. The controller will execute the set of instructions which are stored in the program memory of ROM. The stored instruction is hex code 16-bit wide, first 8-bit as opcode and second 8-bit as operand. Opcode is defined as the type of operation need to be performed on the operand, where operand is just an 8-bit data. Hence operand is the main element in ALU as the type of operation is performed on operand. A lot of time consumption is there for supplying the operand to ALU in each case. A new technique is used to reduce the testing time. For example in ALU test sequence, a seed operand is supplied by the test program in order to speed up the testing process. Here seed operand is just a predefined 8-bit data on which all the Arithmetic and logic operation are going to be performed. After each operation it will get modified for next operation.By designing seed and test routine properly a fault can be detected. Similarly testing of other modules is performed. Considerable time is spent on testing the memory, as it requires read and write operations at every address location. Testing time of memory depends on its size. We have designed RAM and ROM to make memory testing easy, as our main aim is to test embedded controller and not memory. A lot of work has been going on Memory testing [15]. By storing some instructions in the program memory (ROM) of the embedded controller and by executing those instructions, testing of embedded controller is performed. Here the knowledge of instruction set is required for making test program.

Table.1.Testing times for various modules

| Module | Execution Time (cycles) |
|---|---|
| ALU | 83 |
| Program Counter | 64 |
| Instruction Register | 28 |
| Memory (8x8) | 197 |
| Control Unit | 136 |
| Embedded Controller | 245 |

Table 1 shows the time used for testing the various modules. Testing time is calculated by calculating the number of clock cycles [16] taken by the testing program multiplied by the time

58

taken to execute one clock cycle. Time taken depends upon the frequency at which microcontroller is working. We have used ATmega8515, the maximum frequency of 16Mhz up to which it works. The speed can be further increased by using microcontroller of higher frequency i.e. PIC microcontroller can execute up to 40Mhz.

Table2 shows the comparison of area and the delay of the embedded controller with the TMR technique implemented on the two main blocks i.e. ALU and control unit. The slices get almost doubled due to triplication of the modules. Due to small size and low cost of the semiconductor components, the hardware duplicity can be accepted at VLSI level. The TMR technique is very useful in fault masking and detecting the faults in the design. The faults are injected successfully by using fault injection techniques [17], [18] and upon diagnosis; faults are detected hence ensuring the fault coverage. There is little difference in the minimum period shown in Table.2 which can be accepted. The testing time for both the techniques viz. with TMR and without TMR is same excepting the time required for fault injection. The time required for fault injection is very less for a circuit and can be neglected for the testing process.

Table.2. Comparison with and without TMR technique

| Parameters | Without TMR | With TMR |
| --- | --- | --- |
| Utilized Area (Slices) | 112 | 201 |
| Minimum Period (ns) | 7.338 | 7.540 |

## 5. CONCLUSION

Our testing approach successfully tests the Digital IP-core. Upon inserting fault in the design, diagnosis approach detects and locates the faults by using TMR and macros. The proposed approach has been implemented on 8-bit embedded controller as a case study. However it can also be implemented on any other Digital IP-core. Our approach needs no extra test pattern generator and output response analyzer. Since the testing can be performed by using microcontroller and FPGA only, our testing approach is cost efficient, and occupies less area as there is no need for big testers. As FPGA are easily available, the testing can be performed by prototyping an IP-core, faster than hardware emulators. The testing approach is also flexible because of FPGA as it is field programmable and can be reconfigured.Since most of the testing process is based on software i.e. testing programs of microcontroller, it can be reconfigured.

## Authors Biography

**Balwinder Singh** has obtained his Bachelorof Technology degree from National Institute of Technology, Jalandhar and Master of Technology degree from University Centre for Inst. & Microelectronics (UCIM), Punjab University, Chandigah in 2002 and 2004 respectively. He is currently serving as Senior Engineer in Centre for Development of Advanced Computing (CDAC), Mohali andis a part of the teaching faculty and also pursuing Phd from GNDU Amritsar. He has 6+ years of teaching experience to both undergraduate and postgraduate students.Singh has published three books and many papers in the International & National Journaland Conferences.His current interest includes Genetic algorithms,Low Power techniques, VLSI Design & Testing, and System on Chip.

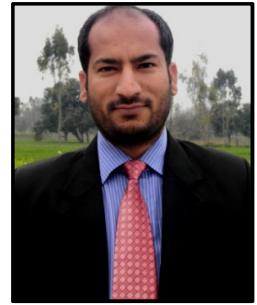

**Amandeep Singh** received the B.Tech. (Electronics and Communication Engineering) degree fromthe Beant College of Engineering and Technology, Gurdaspur affiliated to Punjab Technical University, Jalandhar in 2006, and presently he is doing M.Tech. (VLSI design) degree from Centre for Development of Advanced Computing (CDAC), Mohali and working on his thesis work. His area of interest is Embedded Systems, VLSI Design and Testing, System on Chip, MEMS etc.

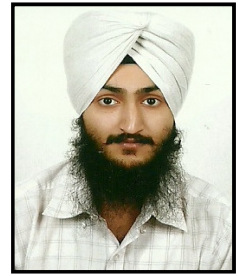